# Superconductivity arising from pressure induced Lifshitz transition in Rb$_2$Pd$_3$Se$_4$ with kagome lattice

Qing Li[†], Yuxiang Wu[†], Xinwei Fan[†], Ying-Jie Zhang, Xiyu Zhu, Zhengyan Zhu, Yiwen Li, Hai-Hu Wen*

National Laboratory of Solid State Microstructures and Department of Physics, Center for Superconducting Physics and Materials, Collaborative Innovation Center for Advanced Microstructures, Nanjing University, Nanjing 210093, China

[†]These authors contributed equally to this work. *e-mail: hhwen@nju.edu.cn

**According to the Bardeen-Cooper-Schrieffer (BCS) theory, superconductivity usually needs well defined Fermi surface(s) with strong electron-phonon coupling and moderate quasiparticle density of states (DOS). A kagome lattice can host flat bands and topological Dirac bands; meanwhile, due to the parallel Fermi surfaces and the saddle points, many interesting orders are expected. Here, we report the observation of superconductivity by pressurizing a kagome compound Rb$_2$Pd$_3$Se$_4$ using a DAC anvil. The parent compound shows an insulating behavior; however, it gradually becomes metallic and turns to a superconducting state when a high pressure is applied. High pressure synchrotron measurements show that there is no structural transition occurring during this transition. The density-functional-theory (DFT) calculations illustrate that the insulating behavior of the parent phase is due to the crystalline field splitting of the partial Pd-4$d$ $t_{2g}$ bands and the Se-derivative 4$p$-band. However, the threshold of metallicity and superconductivity are reached when the Lifshitz transition occurs, leading to the emergence of tiny Fermi surface at $\Gamma$ point. Our results point to an unconventional superconductivity and shed new light on understanding the electronic evolution of a kagome material.**

Layered two-dimensional (2D) van-der-Waals materials have attracted flurry and broad research interests due to their tunable properties by pressure/strain, and thus possess wide application prospects in devices [1, 2]. In particular, the presence of kagome sublattice in such systems may drive the materials into many exotic ground states, such as the quantum spin liquid (QSL) state [3, 4], topological nontrivial electronic state [5, 6], superconductivity [7-9], and so on. For example, the dispersionless flat-bands and the linearly dispersive energy bands have been reported in some 2D kagome metals with 3$d$ transition elements, like $Fe_3Sn_2$, $Co_3Sn_2S_2$, *etc*. [10-13] Although it is rare to find superconductivity in materials with a kagome lattice, it still exists in a few compounds，such as the intermetallic "132" phase $Ba_{2/3}Pt_3B_2$, $LaRu_3Si_2$, *etc*. [14-16] Recently, a new family of vanadium-based layered kagome material $AV_3Sb_5$ ($A$ = K, Rb, and Cs) has attracted significant attention, in which the charge density wave (CDW) order and superconductivity were observed [17-23]. Meanwhile, an anomalous Hall effect was also detected, which was explained as the consequence of some orbital current orders with time reversal symmetry breaking [24-26], although some observations are still at odd with this statement [27]. In addition, by applying pressure or chemical doping, the $AV_3Sb_5$ family shows an unusual competition between superconductivity and CDW order [28-31].

Besides the kagome metals, the kagome insulators have been regarded as the frustrated quantum magnets in the past decades, for example, the quantum spin liquid candidate $ZnCu_3(OH)_6Cl_2$ [32, 33]. We have noticed that there are a series of insulating ternary transition metal chalcogenide $A_2M_3X_4$ ($A$ = K, Rb, Cs; $M$ = Ni, Pd, Pt; $X$ = S, Se) with kagome nets of $M$ ions [34-39]. The crystal structure of $A_2M_3X_4$ was usually reported to be in orthorhombic symmetry with a space group of *Fmmm* (No. 69) or *Fddd* (No. 70) [40, 41]. This family has a quasi-2D layered structure which is constructed by transition metal elements forming kagome nets together with chalcogenide atoms. Among the family, $Rb_2Ni_3S_4$ is the one that has attracted much attention [37-39, 42-45]. The DFT calculation proves that $Rb_2Ni_3S_4$ is a band insulator with substantial flat bands immediately below the Fermi Level [39]. Photoemission spectroscopy study reveals that it is a moderately correlated system with strong hybridization between Ni-3$d$ and S-3$p$ orbitals [44], which is consistent with the band structure calculations [39]. The magnetic and electric transport measurements show that Rb$_2$Ni$_3$S$_4$ is a non-magnetic insulator due to the low spin state of Ni$^{2+}$ in the square planar environment of (S$^{2-}$)$_4$ ligands [37-39, 43]. And the NMR study further confirms that magnetism is absent in the title material [45]. However, up to now, studies in depth on the $A_2M_3X_4$ family are still very limited. High pressure is a clean and controllable method to tune basic structural and electronic properties without changing the chemistry. For layered 2D materials with weak interlayer coupling, pressure may play a more important role. For example, pressure induced insulator-metal transition and superconductivity have been observed in different types of 2D materials, such as FePSe$_3$, CuP$_2$Se, *etc.* [46-49]

In order to find superconductivity in this type of kagome materials, we first performed high pressure resistance measurements on Rb$_2$Ni$_3$S$_4$. Although an insulator-metal transition occurs under high pressures, no superconductivity was observed under high pressure up to 60 GPa. Then we turned to Rb$_2$Pd$_3$Se$_4$, in which the insulating behavior is much weaker at ambient pressure. We report the pressure induced metallization and superconductivity in the title layered compound with kagome lattice. At ambient pressure, Rb$_2$Pd$_3$Se$_4$ is a band insulator with Pd$^{2+}$ in a low spin state. The gap opening around Fermi level may be ascribed to the crystalline field splitting of the partial Pd-4$d$ bands and the Se-4$p$ bands as depicted by the band structure calculations. By applying pressure, the insulating behavior is gradually suppressed, which is accompanied by a monotonic reduction of the lattice parameters. Even though there is no structural phase transition during compression, surprisingly, superconductivity occurs around the boundary of insulator-to-metal transition and a maximum critical transition temperature ($T_C$) of about 3.6 K at 50.6 GPa is obtained. Through DFT calculations, we find that the applied pressure can push the conduction and valence bands closer, and eventually produces a small Fermi surface at $\Gamma$ point when the superconductivity emerges. Thus, we attribute the appearance of superconductivity in pressurized Rb$_2$Pd$_3$Se$_4$ to the Lifshitz transition of the band structure.

## Results

**Sample characterization and physical properties of Rb$_2$Pd$_3$Se$_4$ at ambient pressure.**

According to literatures [40, 41], Rb$_2$Pd$_3$Se$_4$ has an orthorhombic symmetry and weak interlayer van der Waals interactions along *b* axis. Figure 1(a) shows the crystal structure of Rb$_2$Pd$_3$Se$_4$ viewed along *c* axis. Here Rb, Pd and Se atoms are represented by the orange, navy, and green spheres, respectively. The structure can be described as the stacking of infinite layer blocks [Pd$_3$Se$_4$]$^{2-}$ and alkali metal sheets in the atomic layer sequence, -Rb-Se-Pd-Se-Rb-, along the *b* axis. Due to the unusual stacking form, -A-B-A-B-, between layers, the views of just one single layer along the *b*- axis are given in Fig.1 (b). A perfect kagome net of palladium can be seen. The kagome layers are sandwiched between the graphene-like honeycomb networks of Se, and the slabs are separated by the simple hexagonal nets of Rb [Fig.1 (c)]. The powder X-ray diffraction (XRD) pattern and corresponding Rietveld fitting curve of polycrystalline sample are given in Fig.1 (d) and Supplementary Table S1. The Rietveld refinement was carried out with the TOPAS 4.2 software [50]. All the observed peaks can be indexed by an orthorhombic structure with space group: *Fddd* (No.70) and no impurity phase was detected in our measurements. The lattice parameters derived from the Rietveld fitting are *a* = 11.0745(3) Å, *b* = 27.0938(5) Å, and *c* = 6.3554(4) Å, in good agreement with the previous report [40]. We also present a typical single crystal XRD pattern of Rb$_2$Pd$_3$Se$_4$ in Fig.1 (e). Only (0 *l* 0) Bragg peaks can be observed, indicating that the exposed surface is *ac*-plane. The inset of Fig.1 (e) shows the image of an as-grown Rb$_2$Pd$_3$Se$_4$ crystal, and plate-like morphology is formed with an average size of 3 × 3 ×0.5 mm$^3$. Figures1 (f) and (g) present the typical micro-morphology and the energy dispersive spectroscopy (EDS) of the polycrystalline sample. The composition of the compound, determined by EDS measurement, is consistent with the stoichiometry in $Rb_2Pd_3Se_4$ with a small number of Rb vacancies. We also carried out the composition analysis of single crystals and present the data in Supplementary Note 1 and Fig. S1.

Since there was no detailed investigation on $Rb_2Pd_3Se_4$ before, we first conducted a systematic study on the physical properties of $Rb_2Pd_3Se_4$ at ambient pressure. Figures 2(a) and (b) show the magnetic properties of $Rb_2Pd_3Se_4$. If the $Pd^{2+}$ in the kagome nets is magnetic, the strong spin frustration originated from the geometrical restriction may be expected. The temperature dependence of magnetic susceptibility measured at 1 T shows paramagnetic behavior with a Curie like tail at low temperatures. No magnetic order is observed in the whole measured temperature range. The Curie-Weiss (C-W) law is used to fit the $\chi$ - $T$ curve below 40 K by the equation: $\chi = \chi_0 + \frac{C}{T+T_\theta}$, and the fitting yields $\chi_0$ = 8.9×10$^{-4}$ emu mol$^{-1}$, $T_\theta$ = 1.02 K and $C$ = 0.0019 emu K mol$^{-1}$. By using the formula $C = \mu_0 \mu_{eff}^2 / 3k_B$, we can derive the effective magnetic moment $\mu_{eff}$ = 0.07 $\mu_B$/Pd. The magnetization hysteresis loop (MHL) of $Rb_2Pd_3Se_4$ at 10 K within $\mu_0H$ = ±5 T is given in Fig. 2(b). One can see a roughly linear relation between $M$ and $H$, and the magnetization is unsaturated up to $\mu_0H$ = 5 T. The magnetization measurements indicate that the compound $Rb_2Pd_3Se_4$ is a paramagnetic material with negligible effective magnetic moment.

Figure 2(c) presents the temperature dependence of resistivity for $Rb_2Pd_3Se_4$ at ambient pressure in a semilogarithmic scale. Both polycrystal and single crystal (I // ab-plane) samples of $Rb_2Pd_3Se_4$ exhibit insulating behaviors in the measured temperature range. Below 65 K, the resistance value exceeds the measuring range of the instrument. To understand the conduction mechanism of the present material, we present the resistivity data in Arrhenius plots in Fig. 2 (d). The logarithmic resistivity of both polycrystal and single crystal samples is roughly linear with the reciprocal temperature especially in low temperature region (65-160 K) as indicated by the red lines. The linear relation indicates that the insulating behavior of $Rb_2Pd_3Se_4$ at low temperature region can be well described by the band gap model [$\rho = \rho_0 \cdot e^{T_0/T}$]. A deviation of the linear behavior above 160 K may indicate a possible change of the conduction mechanism, or probably it is attributed to the defect state caused by vacancies of rubidium, like the phenomena observed in $Rb_2Ni_3S_4$ or $CrTe_3$ [39, 51]. The temperature dependent specific heat for $Rb_2Pd_3Se_4$ is shown in Figs. 2 (e) and (f). It is known that the low temperature specific heat usually follows the Debye model: $C/T = \gamma_0 + \beta T^2 + \delta T^4 + ...$, where $\gamma_0$ is the specific heat coefficient, $\beta$ and $\delta$ are the fitting parameters for phonon contributions. Then we fit the experimental data using Debye model and present the results in the inset of Fig. 2(e). The fitting results show that the Somerfield term at zero temperature is negligible ($\gamma_0 = 1\times10^{-14}$ mJ mol$^{-1}$ K$^{-2}$), the $\beta$ and $\delta$ are 5.933 mJ mol$^{-1}$ K$^{-4}$ and 0.014 mJ mol$^{-1}$ K$^{-6}$. Using the obtained value of $\beta = 12\pi^4 N_A Z k_B / 5\theta_D^3$, Avogadro constant $N_A = 6.02\times10^{23}$/mol, $Z = 9$, and $k_B = 1.38\times10^{-23}$ J/K, the Debye temperature $\theta_D$ is estimated to be about 143 K. We also display the low temperature $C/T$ versus $T^2$ curve in Fig. 2(f), where the red line is the linear extrapolation of the low-temperature specific heat data considering the Debye model. It is obvious that no residual specific heat at $T = 0$ K, further confirming the absence of Somerfield term. The $\gamma_0$ reflects the finite quasiparticle DOS at the Fermi energy and it should be zero for a band insulator. Thus, the specific heat and resistivity measurements reveal that the title compound $Rb_2Pd_3Se_4$ should be a band insulator. Considering that the nominal value of outer-shell electronic orbital of $Pd^{2+}$ in $Rb_2Pd_3Se_4$ is $4d^8$, eight electrons should partially fill the outermost five different orbitals for 4d shell. This may result in a metallic ground state, but it is contrary to the insulating behavior observed in the transport measurements. Sometimes strong correlation effect can lead to the insulating ground state in the above situation, like that described by the Mott-Hubbard model [52, 53]. In this case, a sufficiently large strength of the on-site

Coulomb interaction U within a partially occupied level can indeed localize electrons. However, in strong coupling picture, the conducting behavior may be described by the variable range hopping (VRH) model [$\rho = \rho_0 \cdot e^{(T_0/T)^\alpha}$)] [54], which is inconsistent with our experimental observations. Another case is that the crystalline field effect induces the insulating ground state. In this scenario, the crystalline field will lift the degeneracy of the 4$d$ orbitals of Pd and hence open a band gap near the Fermi energy. Combined with the electronic band structure calculations given below, we incline to believe that the insulating behavior in $Rb_2Pd_3Se_4$ comes from the latter one.

**Pressure induced metallicity and superconductivity in $Rb_2Pd_3Se_4$.**

In order to have a comprehensive understanding on the quantum states in $Rb_2Pd_3Se_4$, we applied high pressure to tune its electronic properties. Figure 3(a) shows the temperature dependence of resistance ($R$-$T$ curve) of $Rb_2Pd_3Se_4$ in a semi-logarithmic scale at various pressures up to 50.6 GPa. At low pressures (from 0 GPa to 5 GPa), the $R$ - $T$ curves show typical insulating behavior, and the resistance increases rapidly upon cooling. This insulating behavior is strongly suppressed with increasing pressure up to about 40 GPa and the absolute values of resistance at room temperature also decrease dramatically by about three orders of magnitude, suggesting that the material undergoes a significant electronic structure change under pressure. In Fig. 3(b), the normalized resistance ($R/R_{300K}$) as a function of temperature at selected pressures are plotted. Although the decreasing trend of $R/R_{300K}$ slows down at higher pressures (13.8 GPa $<$ $P$ $<$ 40 GPa), the $R$-$T$ curves maintain typical semiconducting behavior. Interestingly, a sudden drop of resistance at around 2.5-3.8 K is observed when $P$ $>$ 18.6 GPa, indicating a possible superconducting transition. By further increasing pressure to 50.6 GPa, the onset superconducting transition temperatures increase with increasing pressure and the resistance drop becomes more explicit as shown in Fig. 3(c). Note that in the pressure range from 18.6 to 39.9 GPa, although the superconducting transition has been observed, the *R-T* curves of the title compound still show a semiconducting behavior above $T_C$. At the highest pressure (50.6 GPa), zero resistance is reached together with a metallic normal state in the whole temperature range, indicating the successful metallization and emergence of superconductivity in pressurized $Rb_2Pd_3Se_4$. That is to say, the superconductivity in $Rb_2Pd_3Se_4$ is generated in the vicinity of insulator-to-metal transition. Such phenomenon is different from the case in some layered materials with insulating ground state, like $FePSe_3$ [46] and $CrSiTe_3$ [49] where the superconductivity appears away from the insulating state and is usually achieved after a structural transition. However, it is rather similar to the situation in the organic Mott insulator $\kappa$-$(ET)_2(Cu/Ag)_2(CN)_3$ [55, 56] and some correlated materials like $CsFe_{4-x}Se_4$ [57] and $BaFe_2S_3$ [58], in which superconductivity usually emerges near the insulator-to-metal transition.

To further confirm whether the resistance drop is related to a superconducting transition in pressurized $Rb_2Pd_3Se_4$, we measured the low temperature resistance under different magnetic fields at *P* = 50.6 GPa (the highest pressure in our present study). As shown in Fig. 3(d), the resistance drop is gradually suppressed to lower temperature with increasing magnetic field and completely disappears above 1.9 K under 4 T. This confirms that the observed transition is indeed a superconducting transition. We extract the magnetic field (*H*) dependent $T_C$ at 50.6 GPa and plot the data in the inset of Fig. 3(d). Owing to the relatively broader superconducting transition, the $T_C$ is determined by using the criterion of temperature for reaching 90% of normal-state resistance ($R_{6K}$). By fitting the $H_{C2}$ - *T* curve with the Ginzburg−Landau (GL) equation:

$$H_{c2}(T) = \frac{H_{c2}(0)(1-t^2)}{1+t^2} \qquad (1)$$

where *t* = $T/T_C$, the zero-temperature $H_{C2}(0)$ is estimated to be 5.4 T, which is lower than its corresponding Pauli paramagnetic limit ($\mu_0H_p$= 1.84$T_C$ = 6.6 T for $T_C$ = 3.6 K). The obtained $H_{C2}(0)$ can be used to calculate the Ginzburg-Landau coherence length $\xi_{GL}$ by using the formula: $\mu_0H_{c2}(0) = \Phi_0/2\pi\xi_{GL}^2$, where $\Phi_0$ is the quantum flux. The calculated $\xi_{GL}$ for $Rb_2Pd_3Se_4$ at 50.6 GPa is 7.8 nm. We also measured the low temperature resistance by applying different currents, which is shown in Fig. 3(e). The resistance drop also shifts to lower temperatures and the zero-resistance state is lost when the applied current becomes larger. Furthermore, in order to ensure that the observed superconductivity is the intrinsic properties of $Rb_2Pd_3Se_4$ rather than the pressure-induced decomposition of the parent compound or impurity phase, we measured the *R-T* curves with reducing pressure from 50.6 to 3.4 GPa. The results are shown in Supplementary Note 1 and Fig. S2. With the decrease of pressure, the superconductivity in $Rb_2Pd_3Se_4$ gradually disappears, and finally the sample returns to the insulating state again. In addition, we also performed an independent high-pressure resistance measurement on a single crystal sample (Fig. S3). This corroborates the observation of superconductivity in pressurized $Rb_2Pd_3Se_4$. The effect of impurity phases which may generate fragile superconductivity can also be excluded (Supplementary Note 2 and Fig. S4). All the evidence confirms intrinsic superconductivity in $Rb_2Pd_3Se_4$ under pressure and the good repeatability of the high-pressure measurements.

We summarize our experimental results and show a temperature – pressure phase diagram of $Rb_2Pd_3Se_4$ in Fig. 4. As one can see, with increasing pressure, the value of $R_{5K}/R_{300K}$ decreases progressively. And its value is less than 1 above 45 GPa, indicating that the material has entered a metallic state. After 18.6 GPa, superconductivity starts to appear above 2 K, together with a clear slope change of $R_{5K}/R_{300K}$. The slight variation of $T_C$ for different experimental runs may arise from the non-hydrostatic conditions in the high-pressure experiments with solid pressure-transmitting medium. At higher pressures, the $T_C$ increases to ~ 3.6 K and is almost saturated above 50 GPa, giving a possible dome-like superconducting region.

**Crystal structural and electronic structural evolution under high pressure.**

To investigate whether the observed superconductivity in pressurized

Rb$_2$Pd$_3$Se$_4$ is associated with a crystal structural phase transition, we performed *in-situ* high-pressure XRD measurements at room temperature. Figure 5(a) shows the synchrotron XRD patterns collected under different pressures. All the Bragg reflection peaks can be well indexed with its ambient pressure structure (Orthorhombic; space group: *Fddd*), indicating that there is no structural phase transition up to 59.5 GPa. The representative refinement results at 2.4 GPa and 59.5 GPa can be seen in Supplementary Fig. S5. As one can see in Fig. 5(a), all peaks shift to higher angles with increasing pressure, indicating the shrink of the lattice parameters. The pressure dependence of lattice parameters is shown in Fig. 5 (b). These lattice constants show smooth and continuous decrease with increasing pressure. The contraction of *b*-axis [(*b*-*b$_0$*) / *b$_0$* = 23.9% at 59.5 GPa] is relatively larger compared with *a* and *c* (15.4% for *a* and 14.28% for *c*) which could be attributed to the weak van der Waals interactions between the layers along *b* axis. The results indicate that the interlayer interaction becomes more pronounced at high pressure in the title layered compound. We also calculated the *b/a* and *b/c* as a function of pressure and show the results in Supplementary Note 3 and Fig. S6. Both *b/a* and *b/c* decrease with increasing pressure and have a clear slope change at about 20 GPa, in good agreement with the pressure where superconductivity appears. Such correlation suggests that the observed superconductivity may be related to a microstructural change of Rb$_2$Pd$_3$Se$_4$ under high pressure. Figure 5 (c) displays the pressure dependent unit-cell volume and the Birch–Murnaghan (*B-M*) [59] fitting. For material without pressure induced phase transition, the unit cell volume (*V*) as a function of pressure (*P*) can be described with the third-order Birch−Murnaghan formula:

$$P(V) = \frac{3}{2} B_0 \left[ \left(\frac{V_0}{V}\right)^{7/3} - \left(\frac{V_0}{V}\right)^{5/3} \right] \cdot \left\{ 1 + \frac{3}{4}(B' - 4)\left[\left(\frac{V_0}{V}\right)^{2/3} - 1\right]\right\} \qquad (2)$$

where *V$_0$*, *B$_0$*, and *B`* are the unit cell volume at ambient pressure, bulk modulus, and first order derivative of the bulk modulus. The fitting results show *V$_0$* = 1850±7 Å$^3$, *B$_0$* = 28.4 GPa, and *B`* = 4.2(5), respectively. Note that the pressure-induced lattice changes are reversible in our present study. As shown in the upper part of Fig. 5(a), the reflection peaks shift back to its initial position when the pressure is released. This is consistent with the reversibility of superconductivity observed in transport measurements (Fig. S3).

In order to clarify the electronic structure and the origin of superconductivity in Rb$_2$Pd$_3$Se$_4$, we performed the DFT calculations under high pressure and present the results in Fig. 6 and Fig. S7. The results show that the insulating behavior at ambient pressure originates from the crystalline field splitting of the energy bands. As we can see from Fig. 6 (a), in a square planar configuration, each Pd ion is located at the center of four Se ions. In the crystal field of planar (Se$^{2-}$)$_4$ ligands with strong covalency between Pd$^{2+}$ and Se$^{2-}$, the specified energy bands of *d*-orbitals will have a level far above other 4*d*-orbitals of palladium. Figure 6 (b) gives a sketch of local energy configurations of Pd and Se orbitals. The antibonding states composed of the 4*p* orbitals of Se and 4*d*$_{xz}$/*d*$_{yz}$ orbitals of Pd should be the lowest empty band. The calculated band structures and DOSs of Pd and Se also give consistent results as shown in Fig. 6(e) and Figs. S7(a-c). Both the valence band maximum and the conduction band minimum are mainly derived from hybridized Pd-4*d* and Se-4*p* states, while the other states have negligible contributions. Moreover, there are flat bands from *d*-orbitals below $E_F$ due to the kagome lattice of Pd$^{2+}$ [39, 60] and the flat bands can result in a sharp density of states at about -0.3 eV. Theoretical studies predicted that the flat band systems in kagome lattice may cause many exotic many-body phenomena, like superconductivity [7, 8].

We display the band structures and DOSs of Rb$_2$Pd$_3$Se$_4$ under high pressure in Figs. 6(f, g) and Figs. S7(d-f). DFT calculations show that the electronic structure gradually changes under pressure. At $P$ = 27 GPa, the top of the valence band and the bottom of the conduction band touch each other and create a small Fermi surface at $\Gamma$ point (Fig. 6c). The emergence of the new Fermi pocket around Brillouin zone (BZ) center suggests that the system evolves towards a Lifshitz transition [61-63]. With further increasing pressure, the conduction bands and the valence bands are fully merged indicating the realization of a metallic state in $Rb_2Pd_3Se_4$ under pressure. For example, at $P$ = 35 GPa, except for the hole pocket around $\Gamma$ point, a separate electronic pocket appears in the vicinity of the $\Sigma_0$ point (Fig. 6f), and there are two sets of Fermi surfaces in the BZ as shown in Fig. 6d. We also present the projected DOS of $t_{2g}$ orbital in $Pd^{2+}$ of $Rb_2Pd_3Se_4$ on the left panel of Figs. 6(f, g), the DOS shows a narrow dip around the Fermi level. The total DOS of $Rb_2Pd_3Se_4$ is given in Figs. S7(d-f) and the calculated carrier density $n$ is about $10^{18}$ /cm$^3$ at both 27 and 35 GPa. It means that the carrier density of $Rb_2Pd_3Se_4$ under pressure is pretty low. The above band structure calculation results show that the metallization and superconductivity in $Rb_2Pd_3Se_4$ may be closely related to the change of the Fermi surface topology (Lifshitz transition).

## Discussions and Conclusions

After systematic analysis of the data, the major finding of the present work is the pressure induced metallicity and superconductivity in $Rb_2Pd_3Se_4$. Firstly, we want to emphasize that the emergence of superconductivity is reproducible and reversible, and the influence of other impurity phases can be safely excluded (Supplementary Notes 1 and Note2). Thus, the superconductivity observed here is an intrinsic property of the title compound. Then we intend to discuss the origin of superconductivity in pressurized $Rb_2Pd_3Se_4$. Since superconductivity occurs without structural phase transition, we can rule out the possibilities of structural transition induced superconductivity as in FePSe$_3$ [46] and CrSiTe$_3$ [49]. We know that there are slope changes in *b/a* and *b/c* versus pressure curves (Fig. S6), and the anomaly is located right in the pressure region where the superconductivity appears. Such facts indicate that the appearance of superconductivity may be closely related to some microstructural instabilities or enhanced interlayer coupling, similar to the situations in CsV$_3$Sb$_5$ [64]. From the DFT calculations, we observed a band-width-controlled formation of a new Fermi pocket at $\Gamma$ point around 27 GPa.

And then another hole pocket appears around $\Sigma_0$ point at about 35 GPa, and further changes the topology of the Fermi surface. That is, two Lifshitz transitions occur in $Rb_2Pd_3Se_4$ under high pressure. Thus, we may conclude that the origin of superconductivity in $Rb_2Pd_3Se_4$ might be the pressure induced Lifshitz transition through the change of the Fermi surface topology. Lifshitz transition is an electronic topological transition of the Fermi surfaces, which modulates the low energy electronic structure, such as creation or disappearance of a Fermi pocket and formation of a new Fermi surface [61, 62]. The pressure induced Lifshitz transition may have great impacts on the exotic phenomena, such as thermoelectricity and superconductivity [65-67]. We argue that the pressure induced metallicity and superconductivity may be attributed to the appearance of Fermi pocket at $\Gamma$ point around the pressure of 27 GPa. Within the framework of the electron-hole quasi-nesting scenario, the subsequently formed hole pocket at $\Sigma_0$ point around 35 GPa (second Lifshitz transition) may be harmful to superconductivity due to the growth of electron pocket at $\Gamma$ point. Thus, the competition of the electron and hole pockets after 35 GPa may explain the saturation of $T_C$ under higher pressures (Fig. 4).

In this regard, we argue that the pairing mechanism of superconductivity in $Rb_2Pd_3Se_4$ may not fall into the BCS scenario, where the high density of state and well-defined Fermi surface(s) are required. In pressurized $Rb_2Pd_3Se_4$ at the threshold pressure for superconductivity (18.6-30 GPa), the Fermi surface is quite small and the calculated carrier density around $E_F$ is rather low, and the BCS superconductivity may not be achieved. Furthermore, the R-T curves around $T_C$ (Figs. 3c-e) show a superconducting-fluctuation-like behavior. Such phenomenon is sometimes considered to be the feature of preformed Cooper pairs in unconventional superconductivity as in cuprates [68]. From the structural point of view, $Rb_2Pd_3Se_4$ is fairly similar to the recently discovered $AV_3Sb_5$ (A = K, Rb, and Cs) family [17-23], in which a kagome net of V exists. Usually, the kagome lattice will host symmetry-protected Dirac points with nontrivial band topology. We also observed a linearly dispersive energy band and Dirac corn around $\Gamma$ point from the calculated band structure at 35 GPa. Thus, the superconductivity in pressurized $Rb_2Pd_3Se_4$ may also host some exotic properties, like topological superconductivity. These possibilities will be examined in future work.

In summary, we have investigated the high pressure structural and electrical transport properties of $Rb_2Pd_3Se_4$ by combining the synchrotron x-ray diffraction and resistance measurements. The structure at ambient pressure is confirmed to be stable up to 59.5 GPa, and the compound exhibits superconductivity with a maximum critical transition temperature ($T_C$) of about 3.6 K at 50.6 GPa. Through the DFT calculations, we find that the appearance of superconductivity is closely linked to the pressure induced topological (Lifshitz) transition of Fermi surface around $E_F$. Our present findings represent the first experimental observation of superconductivity in layered $A_2M_3X_4$ ($A$ = K, Rb, Cs; $M$ = Ni, Pd, Pt; $X$ = S, Se) family with kagome lattice of transition metal. The observation will stimulate further exploration of superconductivity, nontrivial band topology, and potential novel physics in layered materials with kagome lattice.

## Methods

### I. Sample growth and preparation

The polycrystalline samples of $Rb_2Pd_3Se_4$ were synthesized by a conventional solid-state reaction method. Stoichiometric amount of high purity element Rb (99.9%), Pd (99.9%, Alfa Aesar), and Se (99.99%, Alfa Aesar) were used as starting materials. The raw materials were grounded and sealed in an evacuated quartz tube, heated to 850 °C and maintained at this temperature for 24 hours in a muffle furnace. To improve the sample quality, the sample were pressed into pellets and sintered following the previously described temperature procedure once again. Single crystals of $Rb_2Pd_3Se_4$ were grown by a flux method starting from $Rb_2CO_3$ (99%), Pd (99.9, Alfa Aesar), and Se

(99.99%, Alfa Aesar) in a molar ratio of 6:1:20. The mixture was put in an alumina crucible and heated in a muffle furnace with flowing nitrogen gas. The furnace was first sustained at 600°C for 10 hours to prevent the sublimating of Se and then heated at 0.3°C/min to 850°C, holding at this temperature for 10 hours. After that, it took 75 hours to slowly cool down to 600°C. The furnace was then shut down and the single crystals were obtained.

II. Physical properties measurements

The crystal structure was evaluated via XRD (Bruker, D8 Advance diffractometer) with Cu *K*α radiation (λ = 1.543 Å). The Rietveld refinements [69] were conducted with the TOPAS4.2 software [50]. Scanning electron microscopy (SEM) images and energy dispersive spectrum (EDS) were collected on a Phenom ProX scanning electron microscope with an accelerating voltage of 15 kV. The dc magnetization measurements were performed with a SQUID-VSM-7T (Quantum Design). The specific heat was measured by the thermal-relaxation method with the PPMS-16T (Quantum Design). Electrical resistivity measurement at ambient pressure was measured by employing a standard four-probe method. The resistance data under high-pressure was obtained by using four-probe van der Pauw method with platinum foil as electrodes. Diamond anvil cell (*cryo*DACPPMS, Almax easyLab) with 300μm culets was used to generate the pressure up to 60 GPa. Electrical transport measurements under both ambient and high pressure were carried out on PPMS-9T (Quantum Design) in a temperature range from 2 K to 300 K. *In situ* high pressure XRD measurements were carried out at BL15U1 beamline at the Shanghai Synchrotron Radiation Facility (SSRF). A monochromatic x-ray beam with a wavelength of 0.6199 Å was adopted. The sample pressures in our study were determined via the ruby fluorescence method [70].

III. Theoretical calculations

The first principle calculations were performed within density-functional-theory framework by utilizing the full-potential linearized augmented plane wave code WIEN2k [71, 72]. The cutoff parameter $R_{mt}K_{max}$ = 8 along with the muffin-tin sphere 2.5 a.u., 2.2 a.u. and 2.12 a.u. for Rb, Pd and Se atoms was adopted throughout the calculations. We utilized 8000 k-points in the irreducible Brillouin zone for the self-consistent-field calculations and the optimization for the structures under pressure. A denser k-mesh including 20000 k-points was adopted for the calculations of density of states. Both lattice constants and atomic positions were fully relaxed under pressure with the ambient structure extracted from experimental data and the deviation of the volume of the unit cell between the optimized and experimental structures under pressure is smaller than 3%. Fermi surface was visualized by FermiSurfer [73], and the color indicates the Fermi velocity. The carrier concentration was calculated by BoltzTrap [74].

## Data availability

Source data and all other data that support the findings of this paper and other finding of this study are available from the corresponding author upon reasonable request.

## Acknowledgements

We thank Jiajia Feng for the kind help in the high-pressure XRD measurement. This work was supported by the National Natural Science Foundation of China (Grant No. 12061131001, 52072170 and 11927809), and the Strategic Priority Research Program of Chinese Academy of Sciences (Grant No. XDB25000000). The HP-XRD was performed at the beamline BL15U1, Shanghai Radiation Facility (SSRF).

## Author Contributions

H.-H. W. conceived the study. Q.L., Y. X. W, and X. Y. Z. synthesized and characterized the structure of the samples. Q.L., Z.-Y. Z, Y. W. L, and H.-H. W. measured the physical properties of samples at ambient pressure. Q.L., Y.-J. Z. and H.-H. W. performed the high-pressure resistance measurement. Q.L., Y.-J. Z. and X. Y. Z. measured the synchrotron HP-XRD. X. W. F. did the DFT calculations. Q. L., and H.-H. W. analyzed the data and co-wrote the manuscript with input from all authors.

## Competing interests

The authors declare no competing interests.

Figures and legends

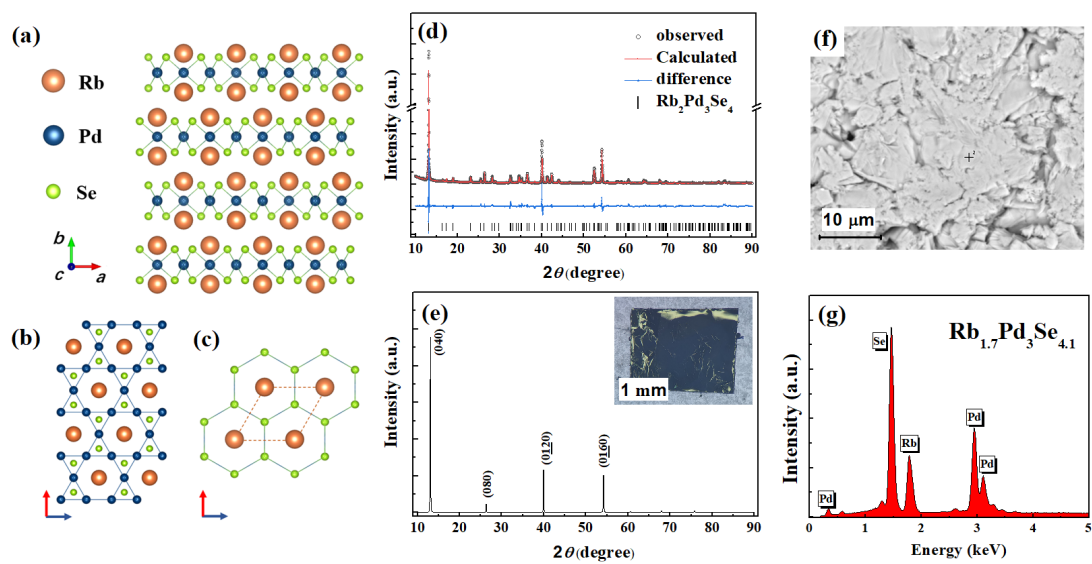

**Figure 1 | Crystal structure and composition characterization of Rb₂Pd₃Se₄. a-c**, Crystal structure of Rb₂Pd₃Se₄ viewed along different axes showing the kagome net of Pd and the layered structure. **d**, Powder X-ray diffraction pattern (circles) and Rietveld fitting curve (red line) of Rb₂Pd₃Se₄. **e**, X-ray diffraction pattern of the Rb₂Pd₃Se₄ single crystal, showing that the largest natural face is (0/0) plane. Inset shows a photo of the Rb₂Pd₃Se₄ single crystal. **f, g**, SEM image and typical energy dispersive spectrum of the polycrystalline sample.

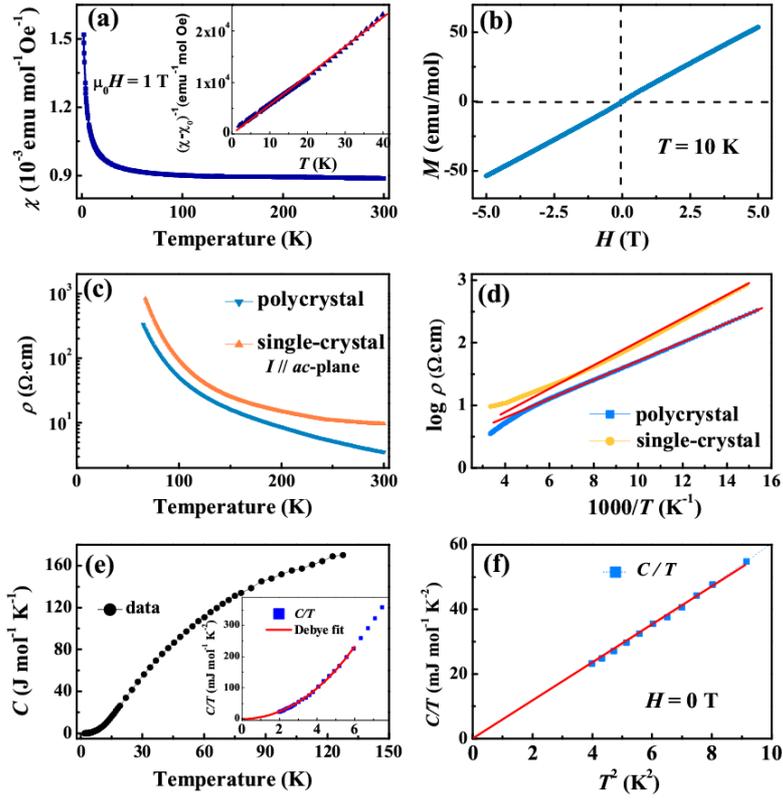

**Figure 2 | Magnetic and transport properties of Rb$_2$Pd$_3$Se$_4$ at ambient pressure. (a),** Temperature dependence of magnetic susceptibility measured with an external field of 1 T. Inset shows the linear relation of $1/(\chi - \chi_0)$ versus $T$ at low-temperature region. **(b),** Magnetization hysteresis loop of Rb$_2$Pd$_3$Se$_4$ at 10 K, shows a linear relation between $M$ and $H$ in the whole measured field range. **(c, d),** Temperature dependence of resistivity curves and the Arrhenius plots (log $\rho$ vs. 1000/$T$) of Rb$_2$Pd$_3$Se$_4$ at ambient pressure. The resistivity data at low temperature region can be well fitted by the band-gap model (linear behavior in Arrhenius plots as shown by the red line). **(e),** Temperature dependent specific heat of Rb$_2$Pd$_3$Se$_4$ at zero field. Inset shows the temperature dependence of the specific heat coefficient $C/T$ at low temperature region, and the red curve is the Debye fit. **(f),** Quadratic temperature dependence of the specific heat coefficient $C/T$ at low temperatures.

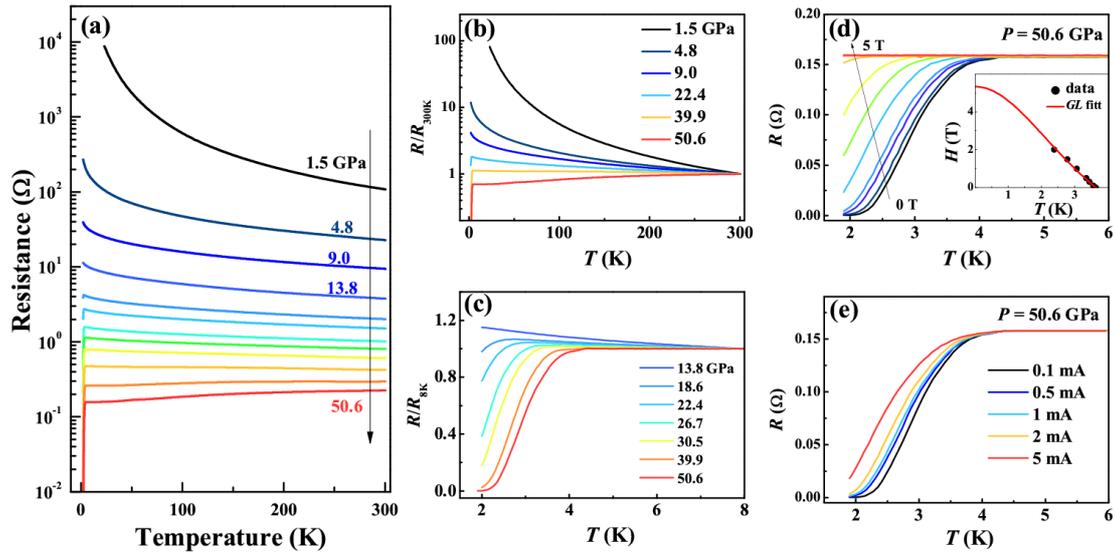

**Figure 3 | The electric transport properties of Rb$_2$Pd$_3$Se$_4$ under pressure.** **(a),** Temperature dependence of resistance (*R-T*) for Rb$_2$Pd$_3$Se$_4$ under various pressures up to 50.6 GPa. **(b, c),** Normalized *R-T* curves of Rb$_2$Pd$_3$Se$_4$ at selected pressures in different temperature ranges. **(d),** The superconducting transition of Rb$_2$Pd$_3$Se$_4$ at 50.6 GPa in various magnetic fields. The inset shows the upper critical field $H_{C2}$ versus *T*, and the red line represents G-L fit. **(e),** The superconducting transition measured with different currents at 50.6 GPa.

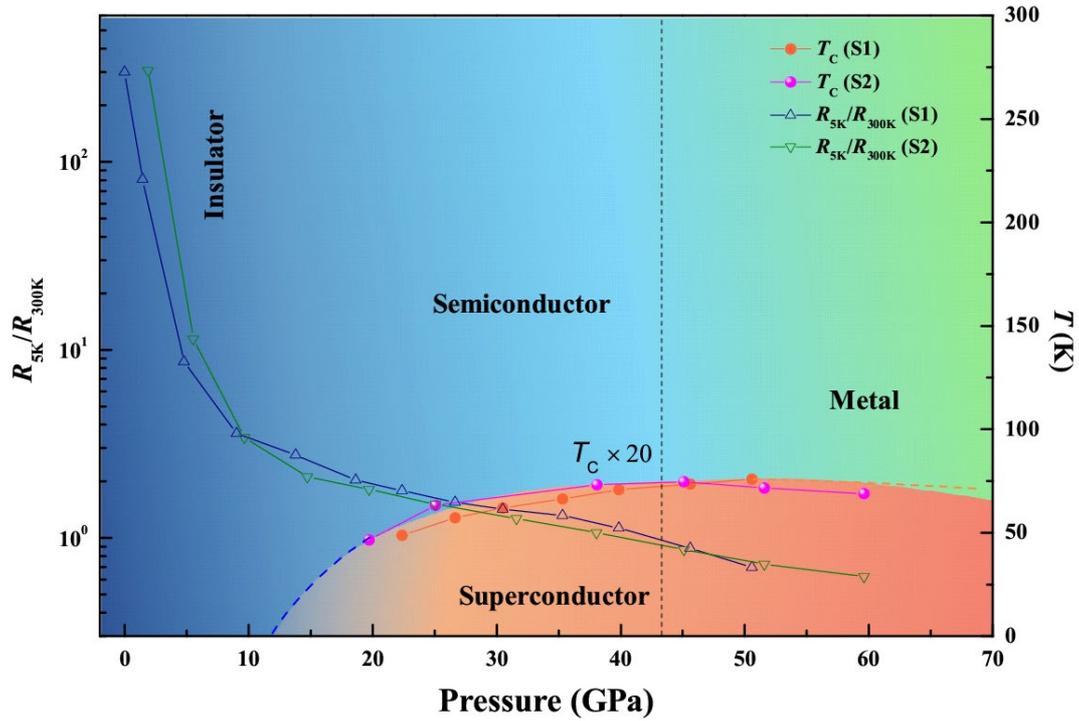

**Figure 4 | Temperature–pressure phase diagram of Rb$_2$Pd$_3$Se$_4$.** The solid circles represent the onset superconducting transition temperatures (20*$T_C$), and the blue and green triangles represent the values of $R_{5K}/R_{300K}$.

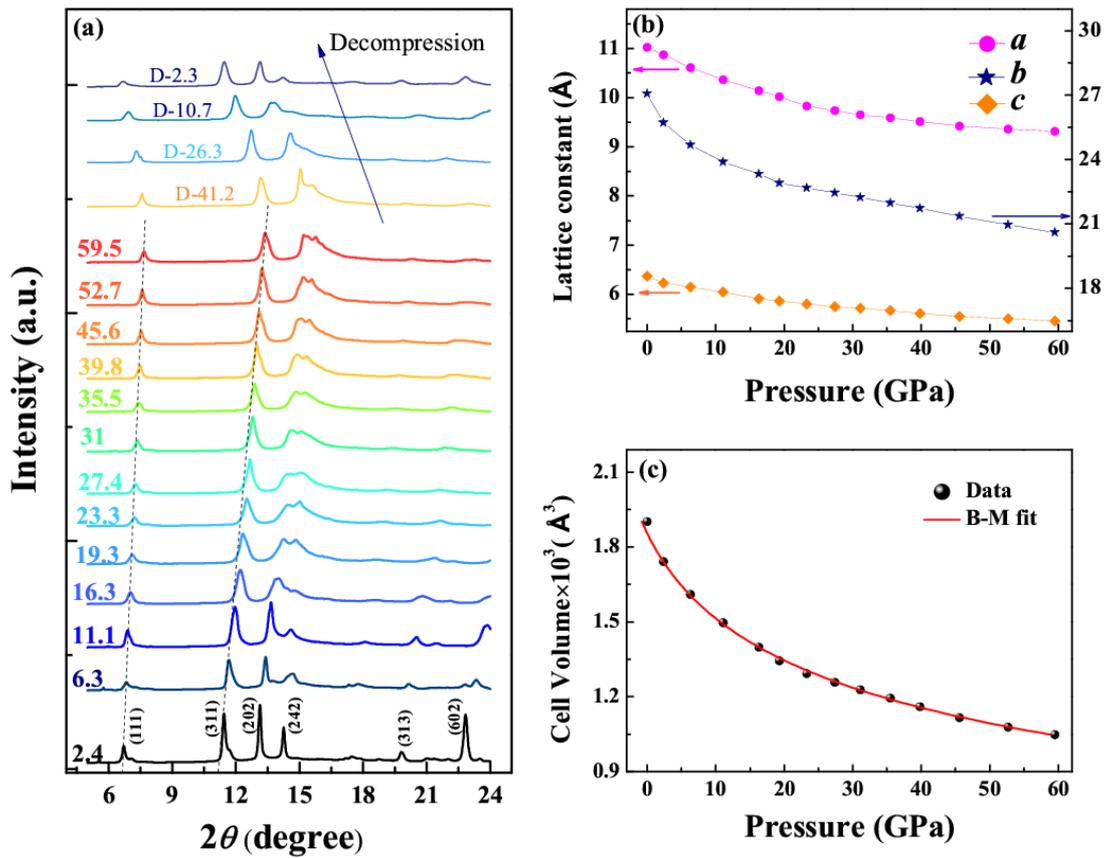

**Figure 5 | High-pressure synchrotron powder XRD analysis of $Rb_2Pd_3Se_4$.** **(a),** X-ray diffraction patterns collected at different pressures up to 59.5 GPa. The typical XRD patterns in decompression process are also given on the top of the panel. **(b),** Pressure dependence of lattice parameters for $Rb_2Pd_3Se_4$ up to 59.5 GPa. **(c),** The derived cell volume as a function of pressure for $Rb_2Pd_3Se_4$. The solid red line is the third-order Birch-Murnaghan fitting line.

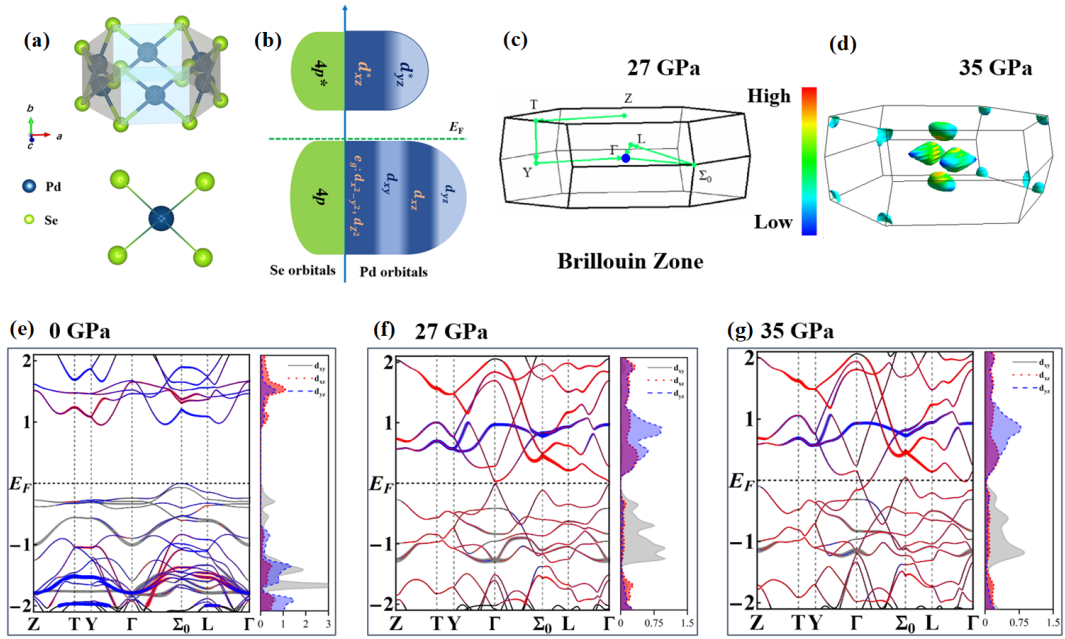

**Figure 6 | Electronic structures of Rb$_2$Pd$_3$Se$_4$ under pressure.**

**(a),** The arrangement of [Pd$_3$Se$_4$]$^{2-}$ blocks and the Pd atoms with a square planar coordination of (Se$^{2-}$)$_4$ ligands. **(b),** The sketch of local energy configurations of Pd and Se orbitals in Rb$_2$Pd$_3$Se$_4$ at ambient pressure. **(c, d),** The Brillouin zone with high symmetry points and Fermi surfaces of Rb$_2$Pd$_3$Se$_4$ under the pressure of 27 and 35 GPa. **(e-g),** The calculated band structures and the projected DOS of $t_{2g}$ orbital of Pd$^{2+}$ at different pressures. The unit of the energy is eV. The size of the red, blue, and gray lines represents the weight of $d_{xz}$, $d_{yz}$, and $d_{xy}$ orbitals of Pd$^{2+}$.